\newcommand{\si}{ \sigma}
\newcommand{\g}{\gamma}
\newcommand{\vq}{\vec{q}}
\newcommand{\vqs}{\vec{q}\,^\prime}
\newcommand{\vl}{\vec{l}}
\newcommand{\vQ}{\vec{Q}}
\newcommand{\vk}{\vec{k}}
\newcommand{\vkq}{\vec{k}-\vec{q}}
\newcommand{\vks}{\vec{k}\,^\prime}
\renewcommand{\d}{\delta}
\newcommand{\D}{\Delta}
\newcommand{\X}{{\chi}}
\newcommand{\bX}{\mbox{\boldmath $\chi $}}
\newcommand{\XR}{{\chi}_{RPA}}
\newcommand{\bXR}{\mbox{\boldmath $\chi $}_{RPA}}

\newcommand{\s}{^{\prime}}

\newcommand{\Gs}{{\bf G}^{\sigma}_0}
\newcommand{\Gms}{{\bf G}^{-\sigma}_0}
\newcommand{\U}{{\bf U}}
\newcommand{\Ggc}{G_{\gamma}^c}
\newcommand{\Ggv}{G_{\gamma}^v}
\newcommand{\Gg}{{\bf G}^{\gamma}}
\newcommand{\p}{{\bf p}}
\newcommand{\sz}{\mbox{\boldmath $\sigma $}_z}

\newcommand{\On}{i\Omega_{n}}
\newcommand{\tOn}{i\tilde{\Omega}_{n}}
\newcommand{\dw}{\Delta\omega}
\newcommand{\w}{\omega}
\newcommand{\wm}{\omega_{\mu}}
\newcommand{\wt}{\omega_{t}}
\renewcommand{\wp}{\omega_{p}}
\newcommand{\wl}{\omega_{l}}
\newcommand{\wn}{\omega_{\nu}}

\newcommand{\pp}{\phantom{+}}

\newcommand{\up}{\uparrow}
\newcommand{\dn}{\downarrow}
\newcommand{\e}{\varepsilon}
\newcommand{\bea}{\begin{eqnarray}}
\newcommand{\eea}{\end{eqnarray}}
\newcommand{\be}{\begin{equation}}
\newcommand{\ee}{\end{equation}}

\documentstyle[aps,prl,preprint,epsfig]{revtex}

\begin{document}
\begin{table}
\end{table}
\begin{figure}
\end{figure}
\newpage
\setcounter{page}{1}
\title{Two-magnon Raman scattering in a spin density wave antiferromagnet}

\author{Friedhelm Sch\"onfeld\footnote{e-mail: fsch@thp.uni-koeln.de, Tel.: 
 0221-470-4208, Fax: 0221-470-5159}, 
 Arno P. Kampf, and Erwin M\"uller-Hartmann  }

\address{
\center{ Institut f\"ur Theoretische Physik, Universit\"at zu K\"oln,}
\center{ Z\"ulpicher Str. 77, D-50937 K\"oln, Germany}}
\date{\today}
\maketitle
\begin{abstract}
We present the results for a model calculation of resonant two-magnon 
Raman scattering in a spin density wave (SDW) antiferromagnet. The resonant 
enhancement of the two-magnon intensity is obtained from a microscopic 
analysis of the photon-magnon coupling vertex. By combining magnon-magnon 
interactions with `triple resonance` phenomena in the vertex function 
the resulting intensity line shape is found to closely resemble the 
measured two-magnon Raman signal in antiferromagnetic cuprates. Both, 
resonant and non-resonant Raman scattering are discussed for the SDW 
antiferromagnet and a comparison is made to the conventional Loudon-Fleury 
theory of two-magnon light scattering.
\end{abstract}

\section{Introduction}
Soon after the discovery of high-T$_c$ superconductivity \cite{Bednorz}
 Raman scattering experiments were performed on the antiferromagnetic (AF) 
 parent compounds $La_2CuO_4$ \cite{Vakmin,Shirane} and $YBa_2Cu_3O_6$ 
\cite{Lyons,Lyons2,Lyons3}.
The analysis of the two-magnon Raman intensity has proven since to be a 
valuable tool for probing the collective magnetic excitations in these 
layered materials.
Common to all antiferromagnetic cuprates is a well defined two-magnon peak 
in the Raman intensity in $B_{1g}$ and a weaker but still significant 
signal in $A_{1g}$ scattering 
geometry at a transferred photon frequency near $3000$ cm$^{-1}$.
The frequency of the two-magnon peak has allowed an estimate for the 
unrenormalized AF exchange coupling between the copper spins in the $CuO_2$ 
planes of about $136 meV$ in $La_2CuO_4$, consistent with results of neutron 
scattering experiments 
for the spin wave velocity \cite{Aeppli} and for the zone boundary magnon 
energy \cite{Hayden,Itoh}. Most of the magnetic properties of AF cuprates are 
well described by model\-ling the undoped $CuO_2$ layers by a spin 
$\frac{1}{2}$ Heisenberg model on a square lattice. Yet, some anomalous 
features of the two magnon intensity profile have remained a puzzle:
both, the asymmetric and broad lineshape in $B_{1g}$ and the appearance of a 
two-magnon signal also in $A_{1g}$ and $B_{2g}$ geometry cannot be obtained 
within the traditional Loudon-Fleury theory for two-magnon Raman scattering
\cite{Fleury,Parkinson,Elliott}.

Furthermore, two-magnon light scattering in AF cuprates is a resonant 
phenomenon and the scattering intensity as well as the line shape 
depend on the incoming photon frequency. 
The Loudon-Fleury theory is in principle a theory for non-resonant Raman 
scattering, assuming a phenomenological coupling of the incoming and 
scattered photons to the localized spins of the antiferromagnet as 
described by the coupling Hamiltonian 
\be 
H_{L-F}= \sum_{<i,j>}(\vec{E}_{inc} \cdot\vec{u}_{ij})
        (\vec{E}_{sc} \cdot\vec{u}_{ij})(\vec{S}_i\cdot \vec{S}_j) 
\label{L-F-H}
\ee
where $\vec{E}_{inc}$ and $\vec{E}_{sc}$ are the electric field vectors 
for the incoming and scattered photons, and $\vec{u}_{ij}$ is a unit vector 
connecting spin sites $i$ and $j$ \cite{Fleury,Parkinson,Elliott}. 
However, light scattering experiments on AF cuprate compounds so far have been 
performed with laser photon frequencies comparable to the charge transfer 
energy gap of these insulating materials \cite{Kampf1,Brenig1}. Therefore, 
photon induced transitions across the insulating energy gap are the natural 
origin for the resonant features of the two-magnon 
signal. A successful theory for resonant two-magnon light scattering must 
for this reason retain the charge degrees of freedom of the electrons.

Along similar lines as in the recent work of Chubukov and Frenkel 
\cite{Chubukov} we perform a model calculation for a spin 
density wave AF which allows to explore the resonant enhancement of the 
two-magnon Raman intensity.
We calculate the scattering intensity using a microscopic description for 
the  photon-electron coupling and for the creation of a magnon  
pair. Final state magnon interactions are included within a diagrammatic 
formulation based on the half-filled single band Hubbard model. 
We show that in this framework the frequency dependence of the photon-magnon 
vertices gives rise to an enhancement of the high energy part of 
the two-magnon spectrum. 
Several experimental features are explained as a consequence of the 
interplay between the two-magnon peak and resonance phenomena of the 
photon-magnon vertex function.

The paper is organized as follows:
In chapter II we start with a brief review of the spin density wave formalism 
for the Hubbard model at half-filling. In chapter III and IV a detailed 
description of our diagrammatic approach is given and the basic coupling 
vertices are calculated. The extension to including final state magnon-magnon 
interactions is presented in chapter V. 
In chapter VI we evaluate results for the non-resonant case and compare them  
 to the conventional Loudon-Fleury theory.
In chapter VII we explore the experimentally relevant resonant case and 
calculate the scattering intensity with resonant electron-magnon vertex 
functions. A discussion of our results and the comparison to the 
experimental data are presented in chapter VIII.

\section{The spin density wave state}
We start from the single band Hubbard model on a square lattice which is 
assumed to describe the low energy 
physics of the $CuO_2$ layers. The Hamiltonian for the Hubbard model 
in standard notation is given by
\be
H=\sum_{{\vk },\si }\e ({\vk }) c^+_{\vk \si} c^{\pp}_{\vk \si } 
   + \frac{U}{N}\sum_{{\vk },\vec{l},{\vq}}c_{\vk\up}^+
   c_{{\vq}-{\vk }\dn}^+c^{\pp}_{{\vq}-\vl\dn}\,c^{\pp}_{\vl\up}
\qquad.
\label{ham}
\ee
Here $c_{\vk \si}^{(+)}$ destroys (creates) an electron with momentum $\vk$ 
and spin $\si = \up , \dn$. $N$ is the number of lattice sites and $U$ is 
the on-site Coulomb repulsion. The 
tight binding dispersion of the square lattice with nearest neighbour 
hopping only is given by 
$\e({\vk})= -4t\g_{\vk}$, where $t$ is the 
hopping amplitude and $\g_{\vk}=[\cos (a{{k}}_x) + \cos (a{{k}}_y)]/2 $. 
Throughout the rest of the paper $t$ and 
the lattice constant $a$ are set to unity.

At half-filling the nesting property $\e(\vk)=-\e(\vk+\vQ)$ with  
$\vQ=(\pi ,\pi)$ leads to an instability of the Fermi sea of non-interacting 
electrons towards a commensurate spin density wave ground state with 
wave vector $\vQ$.
Following the standard procedure as originally outlined by Schrieffer,
Wen, and Zhang \cite{Schrieffer} we introduce the staggered magnetization 
$M$ in the SDW ground state $|\psi \rangle$ by:
\be
 M=\langle\psi |S_{\vQ}^z|\psi \rangle=\langle\psi|\frac{1}{2N}
 \sum_{\vk, \si = \up\dn}\si c_{\vk +\vQ \si}^+ c^{\pp}_{\vk\si}|\psi \rangle
\ee
to linearize the interaction part of the Hamiltonian (\ref{ham}). The 
resulting Hartree-Fock Hamiltonian is diagonalized by the linear transformation
\be
\left( \begin{array}{cc} \g_{\vk \si }^c\\ \g_{\vk \si }^v
\end{array}\right)= {\bf U}(\vk .\si)
\left( \begin{array}{cc} c_{\vk \si  }\\
                         c_{\vk +\vQ \si  }
\end{array}\right)
\quad ,
\quad
{\bf U}(\vk, \si) = \left( \begin{array}{cc} u_{\vk} & \si  v_{\vk}\\ 
                         v_{\vk} & -\si  u_{\vk}\\
\end{array}\right)
\label{trafo}
\ee
which leads to 
\be
H={\sum_{{\vk },\si }}^\prime E({\vk })(\g_{\vk \si}^{c+}
\g_{\vk \si}^{c\pp}-\g_{\vk \si}^{v+}\g_{\vk \si}^{v \pp})
\label{hamg}\quad.
\ee
Here $\g_{\vk \si}^{c(+)},\g_{\vk \si}^{v(+)}$ destroy (create) quasi 
particles in the SDW conduction and valence band, respectively. The 
primed summation is restricted to momenta in the magnetic Brillouin zone 
(MBZ), i.e. to the momenta of the occupied Fermi sea for the non 
interacting system $(\e(\vk) < 0)$, and the quasi particle energy dispersion is 
given by
\be
E({\vk })=
\sqrt{\e^2({\vk })+\Delta^2}\; .
\ee
The SDW energy gap $\D=UM$ between the valence and conduction band is 
determined from the gap equation 
\be
\frac{1}{N}{\sum_{\vk }}^\prime \frac{1}{E({\vk })} = \frac
{1}{U}
\label{SKG}
\ee
and the transformation amplitudes in (\ref{trafo}) are 
given by $u_{\vk}=\sqrt{\frac{1}{2}[1+{\e (\vk )}/{E({\vk })}]}$ 
and $v_{\vk}=\sqrt{\frac{1}{2}[1-{\e (\vk )}/{E({\vk })}]}$. 

Due to magnetic umklapp scattering from the periodic SDW potential the
single particle propagator is no longer momentum diagonal and it is 
conveniently expressed as a $2\times 2$ matrix with respect to the 
momenta $\vq\in$ MBZ 
and $\vq +\vQ$. In this notation the Hartree Fock $c$-particle propagator 
in the SDW state  is written as 
\be
\Gs (\vk ;\w )= \left( \begin{array}{cc}
                      \w + \e (\vk ) & \si \D\\
                      \si\D & \w - \e (\vk ) \end{array}\right)
\frac{1}{\w^2-E^2(\vk) +i\d} \qquad .
\label{G}
\ee
Alternatively we will in subsequent chapters also use the diagonal 
propagator matrix for the SDW conduction and valence band quasi particles. 
The transformation between the $c$ and $\gamma$ representation for the 
propagator matrices reads
\be
\Gs (\vk;\w )= \U^+(\vk ,\si )\Gg(\vk,\w )\U(\vk
,\si )\qquad
\label{gTD2}
\ee
\be
\hspace{-0.5cm}\mbox{with}\hspace{3cm}
\Gg(\vk;\w )=\left( \begin{array}{cc} 
 \Ggc& 0\\ 0 & \Ggv \end{array}\right)
 =\left( \begin{array}{cc} 
\frac{\displaystyle 1}{\displaystyle \w -E(\vk )+i\d } & 0\\ 0 & 
\frac{\displaystyle 1}{\displaystyle \w +E(\vk )-i\d} 
\end{array}\right)\,.
\ee 
The collective spin wave excitations in the SDW state are determined by the 
poles of the frequency Fourier transform of the transverse susceptibilities 
\be
\X^{\si ,-\si}({\vq},{\vqs};t)=\frac{i}{2N}\langle\psi |T S_{\vq}^\si (t)
                                S_{-\vqs}^{-\si }(0) | \psi \rangle \; ,
\ee
where here $\si =\pm$. The local spin raising and lowering operators 
$S_j^\si = S_j^x + \si iS_j^y$ are represented in terms of fermion operators 
by
\be 
S_j^\mu = \frac{1}{2}\sum_{\alpha \beta}c_{\alpha j}^+\si^{\mu}_{\alpha \beta}
          c^{\pp}_{\beta j}\; ,
\ee
where $\si^\mu$ denotes the Pauli matrices with $\mu=x,y$. In the following 
we denote by $\bX(\vq)$ $2\times2$ matrices as in (\ref{G}).

For the calculation of the dynamic transverse susceptibility matrix we 
account for the residual interactions between the quasi particles beyond 
the mean field approximation 
by summing the standard ($RPA$) ladder diagram series \cite{Schrieffer}. 
This leads to the matrix equation:
\be
\bXR^{\si ,-\si }(\vq ;\omega)=\bX_0^{\si ,-\si}(\vq ;\omega)
                 \left[ 1-U\bX_0^{\si ,-\si}(\vq ;\omega )\right]^{-1}\qquad,
\label{chi-RPA}
\ee
with
\bea
 \bX^{\si ,-\si}_0({\vq};\omega )= 
 \frac{1}{N} {\sum_{\vk}}\s&& 
 \left( \begin{array}{cc}
 m^2_{\vk ,\vk +\vq} & \si l_{\vk ,\vk +\vq}m_{\vk ,\vk +\vq}\\
 \si l_{\vk ,\vk +\vq}m_{\vk ,\vk +\vq} & l^2_{\vk ,\vk +\vq}\\
 \end{array}\right) \times \nonumber \\
 & & \left[ \frac{1}{\omega +E(\vk )+E(\vk +\vq )-i\d} \pm 
\frac{1}{\omega -E(\vk )-E(\vk +\vq )+i\d} \right] 
\label{chi_0}
\eea
and the coherence factors 
\be
l_{\vk ,\vk +\vq}=u_{\vk}u_{\vk +\vq}+v_{\vk}v_{\vk +\vq} \quad, \quad
m_{\vk ,\vk +\vq}=u_{\vk}v_{\vk +\vq}+v_{\vk}u_{\vk +\vq}  \qquad.
\label{coh-fact}
\ee
In Eq. (\ref{chi_0}) the upper (lower) sign refers to the off-
diagonal (diagonal) matrix elements, respectively.  
The spin wave dispersion $\w_{sw}(\vq)$ follows from the condition 
$\det[1-U\bX^{+-}(\vq ;\w )]=0 $.
In the strong coupling limit $U\gg t$ the $RPA$ susceptibility matrix 
takes a very  transparent form displaying explicitly the propagating 
spin wave excitations. 
An expansion of $\bXR^{\si ,-\si}$ up to the second order in 
${t}/{U}$ and ${\w}/{U}$ leads to \cite{Brenig}
\be
\bX_{sc}^{\si ,-\si}(\vq ;\w)=\left[ \begin{array}{cc}
 -2J\left(1-\g_{\vq}\right) & \si\w \\
 \si\omega & -2J\left(1+\g_{\vq} \right) \\ \end{array} \right]
 \frac{1}{\w^2-\w_{sw}^2(\vq )+i\d }\; .
\label{chi_sc}
\ee
In this strong coupling limit the spin wave dispersion $\w_{sw}(\vq)=
2J\sqrt{1-(\g_{\vq})^2}$ is identical to the linear spin wave (LSW) 
theory result of the spin $\frac{1}{2}$ Heisenberg antiferromagnet with 
exchange coupling $J=4t^2/U$ \cite{Schrieffer,Brenig,Singh1,Chubukov1}.

In order to describe the coupling of the photons to the electrons we 
consider the Hubbard Hamiltonian in the presence of a weak transverse 
electromagnetic field.
The Coulomb interaction of the Hamiltonian remains  
unchanged but the vector potential $\vec{A}(\vec{r},t)$ of the  
photon field introduces a phase factor $\exp(i\int_l^j\vec{A}(\vec{r},t)\cdot
\vec{dr})$ into the kinetic energy. Expanding 
the kinetic energy part of the Hamiltonian up to second order 
in $\vec{A}$ yields\cite{Shastry}:
\bea
H_{kin}=\sum_{{\vk},\si }\e({\vk}) c_{\vk ,\si}^+
             c^{\pp}_{\vk ,\si} & - &\frac{e}{\hbar c}\sum_{{\vq}}
             \vec{j}({\vq})\cdot\vec{A}(-{\vq})
             \nonumber\\ & + & \frac{e^2}{2\hbar^2c^2}\frac{1}{N}
             \sum_{{\vq}_{1},{\vq}_{2},\alpha,\beta}
          A_{\alpha}(-{\vq}_{1})\tau_{\alpha\beta}({\vq}_{1}+{\vq}_{2})
          A_{\beta}(-{\vq}_{2}).
\label{H-em}
\eea
Here, we have introduced the current density operator $\vec{j}({\vq})$ 
with components 
\be j_{\alpha}({\vq})=\frac{1}{N}\sum_{{\vk},\si }\frac{\partial
                      \varepsilon({\vk})}
                     {\partial k_{\alpha}}c_{{\vk}+{\vq}/{2},\si}^+
                                          c_{{\vk}-{\vq}/{2},\si}^{\pp}
\ee
and the {\it effective} density operator 
\be \tau_{\alpha\beta}({\vq})=\frac{1}{N}\sum_{{\vk},\si }\frac
                        {\partial^2\varepsilon({\vk})}
                        {\partial k_{\alpha}\partial k_{\beta}}
                        c_{{\vk}+{\vq}/{2},\si}^+c_{{\vk}-{\vq}/{2},\si}^{\pp}
\qquad.\ee
In second quantization the components $A_{\alpha}$ of the vector 
potential are expressed in terms of photon annihilation ($a_{{\vq}}$) 
and creation ($a_{{\vq}}^+$) operators by
\be A_{\alpha}({\vq})=\sqrt{\frac{\mbox{h} c^2}{\omega_{{\vq}}\Omega}}\:
      \left(e_{\alpha}^{\pp}a_{-{\vq}}^{\pp}+
            e_{\alpha}^*a^{+}_{{\vq}}\right)\qquad.
\ee
$\vec{e}$ is the polarisation unit vector of the photon and 
$\w_{\vq}=cq$ is the photon frequency. Since the photon wavelength 
for visible light frequencies is large compared to the crystal lattice 
spacing and to all microscopic length scales of the electronic system we 
can safely neglect the photon momenta and henceforth use the $q=0$ limit 
for the photon field. 
 The polarisation vectors of the commonly used experimental 
scattering geometries 
with linearly polarized light are collected in Table 1, 
where $\vec{e}_{i}$ and $\vec{e}_{f}$ denote the polarisation unit vectors
 of the incoming and scattered photon, respectively, with regard to the 
bonds of the $CuO_2$ lattice. 

 The Hamiltonian (\ref{H-em}) describes how the photon field couples to 
the current density operator and determines the bare coupling vertices for 
the common scattering geometries. The second order photon-electron 
coupling to the effective density will be neglected in the following, 
because it contributes 
to the Raman scattering intensity only at high frequencies due to direct 
interband transitions and does not lead to  any resonance phenomena. 
\begin{center}
\begin {tabular}{|c|c|c|}\hline
\hspace{0.5cm} Symmetry \hspace{0.5cm} & $\hspace{1cm}\vec{e}_{i}$
\hspace{1cm}  & $\vec{e}_{f}$\\ \hline\hline
 & & \\
$A_{1g}$ & $\;\frac{1}{\sqrt{2}}(1,1)\;$ & $\;\frac{1}{\sqrt{2}}(1,1)\;$\\
 & & \\
$B_{1g}$ & $\;\frac{1}{\sqrt{2}}(1,1)\;$ & $\;\frac{1}{\sqrt{2}}(1,-1)\;$\\
 & & \\
$B_{2g}$ & $\;(1,0)\;$ & $\;(0,1)\;$\\\hline
\end{tabular}
\newline
\newline
{\small Table 1. Polarisation vectors for the incoming 
$(\vec{e}_i)$ and \newline 
scattered  photon $(\vec{e}_f)$ in the common scattering geometries.\newline} 
\end{center}

\section{Two-magnon Raman scattering}
\begin{figure}
\begin{center}
\epsfig{file=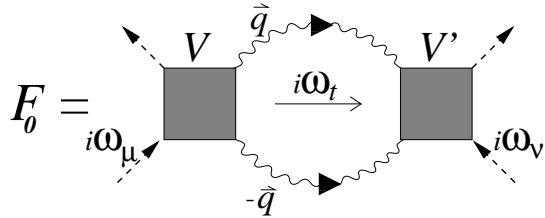, height=3cm}
\end{center}
\caption{\label{general-diag} General diagram for the Raman amplitude of 
two-magnon scattering. The dashed lines represent the incoming and 
outgoing photons with frequencies $i\wm , i\wn$ and the wiggly lines 
the magnon propagators. The vertices $V$ and $V\s$ contain the microscopic 
coupling of the photons to the magnon excitations and $i\wt$ is the 
transferred photon frequency.}
\end{figure}

As an alternative to the Golden Rule analysis of the Raman scattering 
intensity, we follow here a convenient diagrammatic formulation 
\cite{Kawabata}. This method has been 
applied previously for calculating the Raman intensity from two-phonon 
scattering \cite{Klein} or for two spin fluctuation scattering 
in the paramagnetic phase of the Hubbard model \cite{Kampf}. 
Here we extend this technique to a $2\times 2$ matrix formulation for 
the SDW state.
The general diagram for the Raman amplitude in the absence of 
magnon-magnon interactions is shown in Fig. \ref{general-diag}.
The photons couple to the electronic system through a vertex function $V$, 
 creating two magnons with momenta $\vq$ and $-\vq$. 
The left side of the diagram corresponds to the physical process; the 
scattering intensity is then obtained from the symmetrically completed 
diagram by taking diagram cuts 
(see below) - a technique which yields results equivalent to the 
Golden Rule analysis. The virtue of this diagrammatic technique is that 
it allows to select those scattering processes which dominantly contribute 
to the two-magnon Raman signal \cite{Klein}. Here and in the following we use 
the finite temperature Matsubara formalism and after analytic continuation to 
the real frequency axis we 
evaluate our results in the zero temperature limit.

The final state of interest contains two magnons with opposite momenta. 
Therefore the two-magnon scattering intensity can be deduced by taking 
a cut of the diagram for the Raman amplitude across the two
 magnon lines only. After analytic continuation of the photon Matsubara 
frequencies the cut translates into taking the discontinuity 
of $F_0$ across the real frequency axis with respect to the transferred 
photon frequency $i\w_t$, 
\be 
I(\dw) = {\frac{1}{2\pi i}}\left[F_0(i\w_t\rightarrow\dw+i\delta)
                                -F_0(i\w_t\rightarrow\dw-i\delta)\right]
\quad,
\label{cut}
\ee
and  $I(\dw)$ is directly proportional to the scattering intensity.
Explicitly, given the magnon propagators as represented by the transverse 
dynamic spin susceptibilities, 
 and the vertex functions $V$ and $V\s$, which will be calculated in 
chapter IV, the Raman 
amplitude $F_0$ has the following form:
\bea
F_0(i\wt)=\frac{1}{\beta N}\sum_{{\vq},{\vq}\,\s,\si ,i\wp}
 && V_{\si} ({\vq},i\wt,i\wm,i\wp)
    {\X}^{\si ,-\si }({\vq},{\vq}\,\s,i\wp)\;\times
\nonumber \\&&  
   {\X}^{-\si ,\si }({-\vq},-{\vq}\,\s,i\wt-i\wp)
   {V}\s_{\si}({\vq}\,\s,i\wt,i\wn,i\wp)
\label{ampl}
\eea
where $\beta$ is the inverse temperature. 
In order to single out only the two-magnon contribution to $F_0$, 
the residues of the vertex function must be disregarded  
when carrying out the internal frequency sum. Similarly, the cut 
prescription Eq. (\ref{cut}) with respect to $i\w_t$ is restricted 
to the arguments of the susceptibilities. (For details of this technique see Ref. \cite{Klein}.) The omitted high energy contributions correspond to final 
states with particle-hole excitations only. 

The frequency sum in (\ref{ampl}) is conveniently performed by introducing 
the spectral representation for the transverse dynamic spin susceptibility
\be
\bX^{\si , -\si }(\vq ; i\w_p)= -\frac{1}{\pi}\int
\frac{\mbox{Im}\bX^{\si , -\si }(\vq ;\w +i\d)}{i\w_p -\w}d\w
\qquad .
\ee
The analytic continuation of the Matsubara 
frequencies in the vertex function is performed according to the rules 
\cite{Klein,Kampf}
\bea
i\wm&\rightarrow &\omega_i+i\delta 
\label{ana-cont1}\qquad\nonumber\\
i\wn&\rightarrow &-\omega_i+i\delta
\label{ana-cont2}\qquad\nonumber\\
i\wt&\rightarrow &\omega_i-\omega_f=\dw\qquad.
\label{ana-con3} 
\eea
$\w_i$ and $\w_f$ are the frequencies of the incoming and outgoing  
photon, respectively, and $\dw= \w_i - \w_f$ is the photon frequency shift. 
Taking the cut according to Eq. (\ref{cut}) then yields the desired two-magnon 
contribution to the scattering intensity:
\bea
I(\dw)&=&\sum_{{\vq},{\vq\,}\s,\si}\frac{1}{\pi^2N}\int d\omega
V_\si ({\vq},\dw ,\w_i +i\d,-\omega)
\mbox{Im}{\X}_{RPA}^{\si ,-\si }({\vq},{\vq\,\s},-\omega)\times
 \nonumber \\ && 
 \mbox{Im}{\X}_{RPA}^{-\si ,\si }(-{\vq},-{\vq}\,\s,\dw+\omega)
           V^{\s}_{\si}({\vq}\,\s,\dw,-\w_i+i\d ,-\omega)
\left\{n(\omega+\dw)-n(\omega)\right\}.
\label{Iw}
\eea
Here, the Bose distribution function $n(\w )$ is evaluated in the 
$T\rightarrow 0$ limit. Note that $V_\si$ and $V\s_\si$ are in fact the 
same vertex 
functions but differ in the argument for the incoming photon frequency 
due to the analytic continuation rules Eq. (\ref{ana-cont1}). As a 
consequence the primed vertex function $V\s_\si (\vq)$ is equal to the complex 
conjugate of $V_\si(\vq )$.
In order to outline the derivation of the scattering intensity (\ref{Iw}) 
no final state magnon-magnon interactions have been included so far. The 
inclusion of magnon-magnon interactions will be discussed in chapter V.

\section{The vertex function}
\begin{figure}
\begin{center}
\epsfig{file=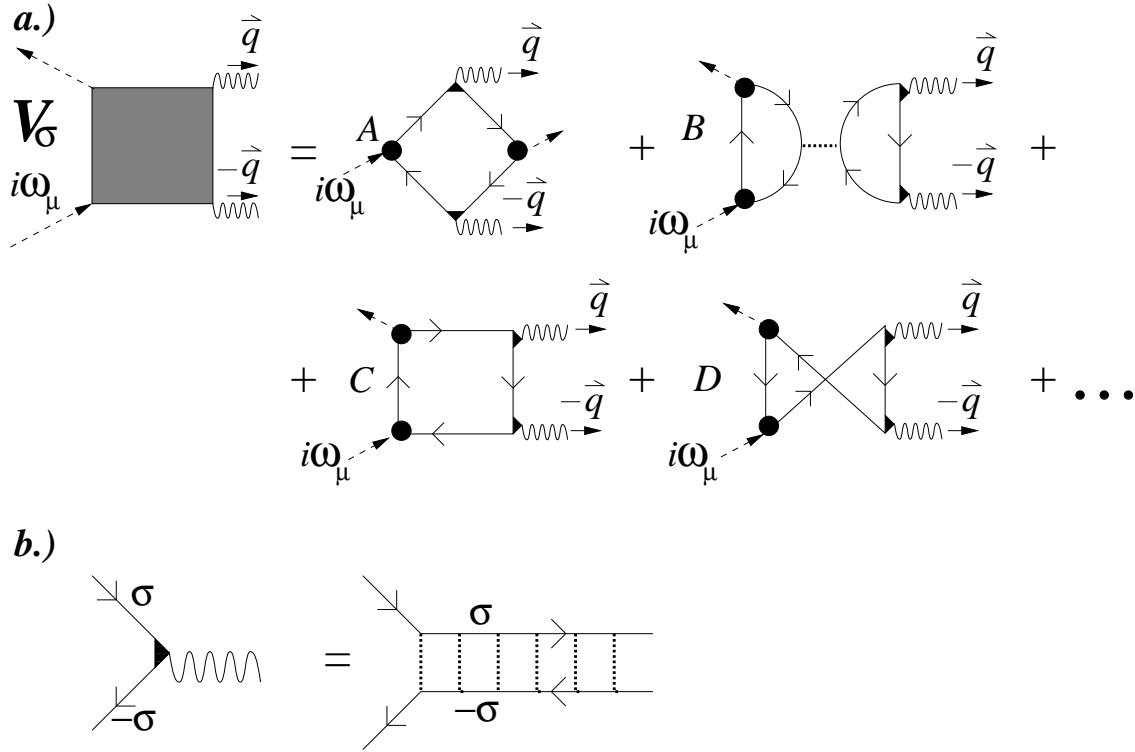, height=10cm}
\end{center}
\caption{\label{vertices}a.) Effective vertex function for the photon-magnon 
 coupling. The solid, dashed, and wiggly lines denote the SDW $c$-fermion, 
 the photon, and the magnon propagator, respectively. 
 The spin labeling of the fermion lines has been omitted, 
 it is explicitly indicated in b.). 
 The solid circle represents the bare photon-electron coupling 
 $\vec{j}\cdot\vec{A}$, 
 the filled triangle represents the electron-magnon coupling where the 
 magnons are contained in the particle-hole ladder series as shown in b.).}
\end{figure}

Now we focus on the photon-magnon vertex function $V_\si$ which depends 
on the incoming light frequency and is thus responsible for the resonant 
behavior of the scattering intensity. The 
electron-photon coupling 
Hamiltonian (\ref{H-em}) determines the bare coupling vertices 
 of the photon's vector potential to the current density. Given the bare 
photon-electron vertices 
the simplest and in the strong coupling limit most relevant contributions to 
the vertex function $V_\si$, which mediates the indirect coupling of the 
photon to the spin wave excitations, are shown by the diagrams in 
Fig. \ref{vertices}.

The algebraic expressions corresponding to the individual diagrams that 
contribute to the photon-magnon vertex function $V_\si$ as shown in Fig.
\ref{vertices} are explicitly listed in the appendix. 
Since the total vertex function $V_\si$ is independent of $\si$  we 
henceforth drop the spin index $\si$.
In addition to the diagrams shown in Fig. \ref{vertices}a  the diagrams 
with reversed direction of the fermion lines 
have to be included as well.  
With Eq. (\ref{Iw}) and the vertex function contributions as listed in 
the appendix the derivation of the Raman intensity in the absence of 
magnon-magnon interactions is complete.

\section{Magnon-Magnon interaction}
\begin{figure}
\unitlength1cm
\begin{picture}(8,8)(0,0)
\put(0.5,4){
\epsfig{file=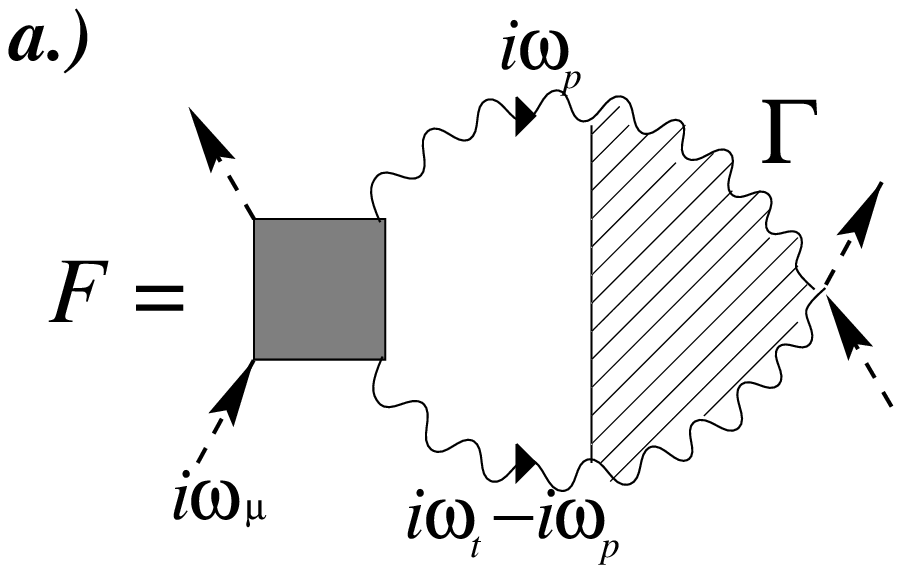,height=3.5cm}
}
\put(0.5,0.2){
\epsfig{file=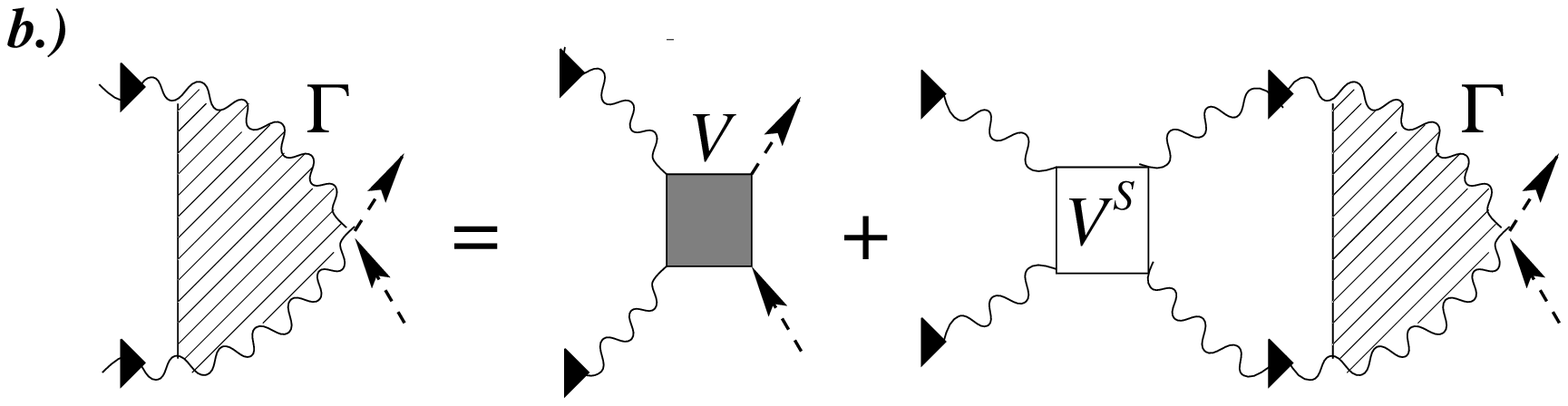,height=3.4cm}
}
\end{picture}
\caption{\label{MMW2}a.) General diagram for the Raman amplitude with 
final state 
magnon-magnon interaction.
b.) Diagram for the renormalized magnon-photon vertex function $\Gamma$ 
which contains only magnon-number conserving scattering processes. The 
solid squares in a.) and b.) represent the bare photon-magnon coupling $V$ 
(see Fig. \ref{vertices}) and $V^S$ in b.) represents the irreducible 
magnon-magnon interaction.}
\end{figure}

The remaining step is to include the effects of final state magnon-magnon 
interactions. 
From the results obtained in the framework of the Heisenberg model it is 
well known that it is crucial to include final state interactions in order 
to avoid a diverging Raman intensity at twice the maximum magnon frequency 
$\dw = 2\w_{sw}^{max}=4J$ resulting from the square root divergence of 
the magnon density of states at the MBZ boundary 
\cite{Fleury,Parkinson,Elliott}. We therefore extend the Raman amplitude 
$F_0$ to allow for repeated magnon-magnon scattering.
This is achieved by replacing a bare vertex $V$ in the Raman amplitude by 
a renormalized magnon-photon vertex function 
$\Gamma$ which contains an infinite series of magnon pair scattering 
processes. 
Diagrammatically this is represented in Fig. \ref{MMW2}a and requires the 
solution of a Bethe-Salpeter like equation for $\Gamma$ (see Fig. \ref{MMW2}b).
The diagrams for the Raman amplitude $F$ and the vertex function $\Gamma$ 
translate into the following equations:
\bea
F(i\wt ,i\wm)=\frac{1}{\beta}\sum_{i\wp}\frac{1}{N}
                               \sum_{\vq_1 ,\vq_1^{\,\prime}}
&V&(\vq_1,i\wm ,i\wt ,i\wp )  \XR^{+-}(\vq_1, \vq_1^{\,\prime} ,i\wp )
\times \nonumber \\&&
  \XR^{-+}(-\vq_1, -\vq_1^{\,\prime} ,i\wt -i\wp )
  \Gamma(\vq_1^{\,\prime} ,i\wm, i\wt, i\wp) \qquad,
\label{gl1} 
\eea
\bea
\Gamma(\vq_1^{\,\prime},i\wm, i\wt, i\wp)&=&
 V(\vq_1^{\,\prime},i\wm ,i\wt ,i\wp )-\frac{1}{\beta}\sum_{i\wl}
 \frac{1}{N}\sum_{\vq_2 ,\vq\,\s_2}
 V^s(\vq_1^{\,\prime} ,\vq_2,i\wp ,i\wt ,i\wl )
 \times\nonumber \\ & &
 \XR^{+-}(\vq_2, \vq_2^{\,\prime} ,i\wl )
 \XR^{-+}(-\vq_2 , -\vq_2^{\,\prime} ,i\wt -i\wl ) 
 \Gamma(\vq_2^{\,\prime},i\wm, i\wt, i\wl) \qquad.
\label{gl2}
\eea

\begin{figure}
\unitlength1cm
\begin{picture}(6,3.5)(0,0)
\put(2,0.5){
\epsfig{file=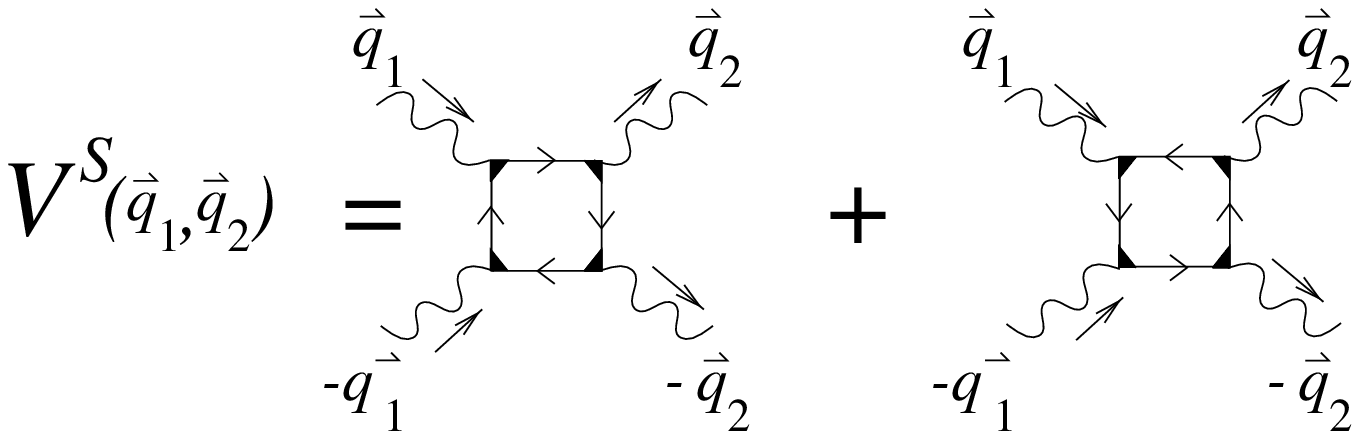,height=3.5cm}}
\end{picture}
\caption{\label{MMW4}Vertices for magnon-magnon scattering. As before 
(see Fig. \ref{vertices}b) the filled triangle corresponds to the 
electron-magnon coupling.}
\end{figure}

For the irreducible magnon-magnon interaction $V^S$ we include only magnon 
number con\-ser\-ving scattering pro\-cesses 
which - at least for the Loudon-Fleury theory of Raman scattering from the 
Heisenberg antiferromagnet - 
have been shown to be the most important \cite{Canali}. For our 
present SDW state based calculation the magnon-magnon interaction vertex 
has to be expressed in terms of the residual Hubbard interaction between 
the fermionic quasi particles.  The simplest diagrams which serve for this 
purpose are shown in Fig. \ref{MMW4} and involve an internal loop with 
4 fermion Green's functions. Fortunately, for our subsequent 
strong coupling evaluation of the two-magnon Raman intensity in $B_{1g}$ 
geometry we have indeed found that these two diagrams give the dominant 
contribution to the magnon-magnon interaction. As we will show below, 
evaluating $V^S$ in the static zero frequency limit the result for the 
$B_{1g}$ intensity can still be obtained analytically for $U\gg t$ in an  
intermediate frequency range despite the mathematical complexity of the 
Raman amplitude $F$ in Eq. (\ref{gl1}).

\section{Non-resonant scattering}
First we analyse the results in the non-resonant case, when the incoming 
photon frequency $\w_i$ is much smaller than the SDW energy gap $2\Delta$. 
In the strong 
coupling limit all expressions take a transparent form and become 
analytically tractable.
In $B_{1g}$ symmetry the rhomb-shaped vertex diagram $A$ in Fig. 
\ref{vertices} gives the only contribution to the total vertex function; 
the diagrams $B,C,$ and $D$ vanish identically to leading order in $t/U$. 
Explicitly we find 
\be
V_A^{sc}(\vq)= -f(\w_i) J \g_{\vq}^d \quad ,
\label{VA_sc}
\ee
\be 
\hspace{-2.8cm}\mbox{with}\qquad
\gamma_{\vq}^d=\frac{1}{2} \left[ \cos({q}_x)-\cos({q}_y)\right] \qquad
\mbox{and}\qquad f(\w_i)\propto \frac {\D^2}{\w_i^2-4\D^2}\; .
\label{gd}
\ee
This simple structure of the photon-magnon vertex function leads immediately 
to a simple expression for the scattering intensity (\ref{Iw}) 
in $B_{1g}$ geometry without final state magnon-magnon interactions: 
\be
I_{B_{1g}}(\dw)\propto 
\frac{\Delta ^4}{(\w_i^2-4\Delta^2)^2}{\sum_{\vq}}\s
\frac{J^2 (\gamma_{\vq}^d)^2}{1-\gamma_{\vq}^2}\delta (\dw - 2\w_{sw}(\vq))\; .
\label{Iw_sc}
\ee
As expected the neglect of  magnon-magnon interactions results 
in a logarithmic divergence of the $B_{1g}$ intensity at twice the 
maximum spin wave frequency $4J$.

In $A_{1g}$ symmetry all 4 diagrams in Fig. \ref{vertices} for the 
vertex function give non-vanishing contributions, 
but due to a perfect cancellation of the different terms the total vertex 
function and therefore the scattering intensity vanish to leading order 
in $t/U$. This cancellation has been previously noted by Chubukov and Frenkel 
\cite{Chubukov}. 
Also the intensity in $B_{2g}$ symmetry is zero in the strong 
coupling limit. For weak and intermediate values of $U/t$, however, the 
two-magnon intensity is 
finite in all three symmetry channels.

The logarithmically diverging scattering intensity in $B_{1g}$ 
(\ref{Iw_sc}) and the vanishing intensities in $A_{1g}$ and $B_{2g}$ 
symmetries for $U\gg t$ are results one also obtains using the effective 
Loudon-Fleury spin-photon coupling Hamiltonian in the framework 
of the spin $\frac{1}{2}$ Heisenberg model. In the non-resonant region, 
where $(2\D -\w_i )\gg J$, and in the strong coupling limit the assumption 
of localized spins is valid and the Loudon-Fleury approach is expected 
to yield the correct description \cite{Chubukov,Shastry}. 
Since we have demonstrated the equivalence between the two approaches at
least in the absence of magnon-magnon interactions,
 the correct selection of the photon-magnon vertex diagrams in 
Fig. \ref{vertices} is a posteriori 
verified.

In order to take magnon-magnon interactions into account, we evaluate
the scattering vertex $V^S$ (see Fig. \ref{MMW4}) in the static limit
for $U\gg t$. In $B_{1g}$ geometry the only relevant contribution to
$V^S$ is given by 
\be V^S_{B_{1g}}(\vq_1,\vq_2 )= -12J\,
\g_{\vq_1 - \vq_2 } \qquad .  \ee 
We decompose $\g_{\vq_1 - \vq_2}$ as 
\be \g_{\vq_1 - \vq_2 }
= \g_{\vq_1}\g_{\vq_2} +
\g^d_{\vq_1}\g^d_{\vq_2} +
\g^{p1}_{\vq_1}\g^{p1}_{\vq_2} +
\g^{p2}_{\vq_1}\g^{p2}_{\vq_2}\quad , 
\label{decomp}
\ee 
where 
\be
\g^{p1}_{\vq} = \frac{1}{2}[\sin(q_x)+\sin(q_y)] \quad\mbox{and}\quad
\g^{p2}_{\vq} = \frac{1}{2}[\sin(q_x)-\sin(q_y)]\,. 
\ee 
Due to the specific momentum dependence of the relevant photon-magnon vertex, 
Eq. (\ref{VA_sc}), only the second term in Eq. (\ref{decomp}) contributes, 
because in performing the momentum sum in the vertex Eq. (\ref{gl2}) all
those parts of $V^S_{B_{1g}}$ which are orthogonal to $\gamma_{\vq}^d$ 
give a vanishing contribution ($\sum_{\vq}\g_{\vq}^d\g_{\vq} =
\sum_{\vq}\g_{\vq}^d\g_{\vq}^{p1}=\sum_{\vq}\g_{\vq}^d\g_{\vq}^{p2}=0$). 
Therefore, the effective remaining magnon-magnon interaction
$V^S_{eff}$ in $B_{1g}$ geometry is given by 
\be V^S_{eff}(\vq_1 ,\vq_2 )= -12 \,
J\,\g^d_{\vq_1}\g^d_{\vq_2} \qquad.
\label{Vs}
\ee
Introducing the function 
\bea
L(i\wt )=\frac{1}{\beta}\sum_{i\wp}\frac{1}{N}\sum_{\vq ,\vq^{\,\prime}}
\gamma_{\vq}^d
\X_{SC}^{+-}(\vq, \vqs ,i\wp )
\X_{SC}^{-+}(-\vq, -\vqs ,i\wt -i\wp ) \gamma_{\vq^{\,\prime}}^d\quad,
\label{L}
\eea
and solving Eqs. (\ref{gl1}) and (\ref{gl2}) with 
$V^S=V^S_{eff}(\vq\,\s_1 ,\vq_2)$ the $B_{1g}$ Raman amplitude takes the form
\be
 F_{B_{1g}}(i\wt )=f^2(\w_i)\frac{J^2 L(i\wt )}{1-12 J L(i\wt )}
   \qquad.\hspace{1cm}
\label{gl4}
\ee
Finally, applying the cut prescription Eq. (\ref{cut}) yields the result for 
the non-resonant scattering intensity in $B_{1g}$ geometry 
\be
I_{B_{1g}}(\dw)=\frac{f^2(\w_i)}{\pi}\mbox{Im}\left\{\frac{L(\dw +i\delta)}
{1-12J\,L(\dw +i\delta)}\right\} \qquad.
\label{Iw_WW}
\ee

A comment is in order regarding the calculation of $L(i\omega_t)$. We note 
that the denominator of 
$\bX_{sc}^{+-}(\vq ;\w)$ in Eq. (\ref{chi_sc}) contains two poles 
at $\w=\pm \w_{sw}(\vq)$ 
corresponding  to forward and backward propagating spin wave excitations. 
The vertices in Fig. \ref{MMW4}, however, account only for magnon number 
conserving scattering processes. For a consistent evaluation of the Raman  
amplitude it is required  to retain only the forward propagating 
parts of the dynamic susceptibilities in Eq. (\ref{L}) for $L(i\w_t)$
\cite{Chubukov3}. 

\begin{figure}
\begin{center}
\epsfig{file=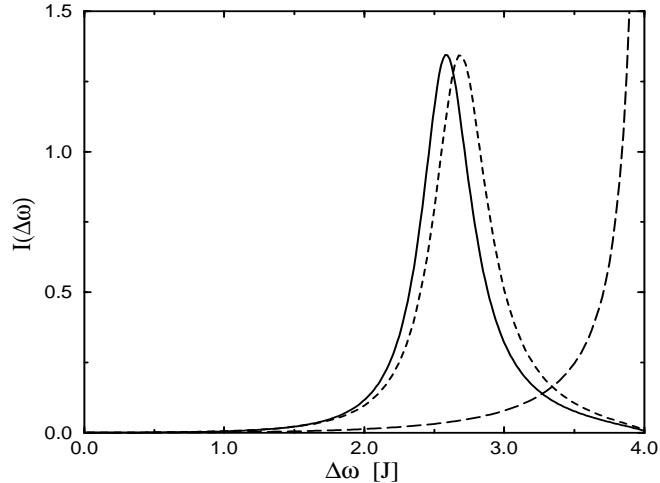,height=7.5cm, width=10cm}
\end{center}
\caption{\label{Erg_sc}Raman intensity in arbitrary units in the 
non-resonant case in the strong coupling limit with (solid line) and 
without (long dashed line) final state magnon-magnon interactions.   
The short dashed line shows the result of the Loudon-Fleury theory with 
magnon-magnon interactions included.
The transferred photon frequency $\dw$ is given in units of $J=4t^2/U$.}
\end{figure}

With this restriction for the evaluation of $L(i\w_t)$ the resulting $B_{1g}$
 scattering intensity is shown in Fig. \ref{Erg_sc} (solid line). 

As expected, the inclusion of magnon-magnon interactions removes the 
divergency of the intensity for zone boundary magnon pairs and leads to 
a single almost symmetric two-magnon peak around $\dw \cong 2.6J$. 
For comparison we have also shown the result of the Loudon-Fleury 
theory in Fig. \ref{Erg_sc} obtained from a spin wave theory analysis 
to order $(1/S)^2$. We note that an analogous but much more tedious 
expansion in the framework of the half-filled Hubbard model has been 
performed in Ref. \cite{{Chubukov2}} by extending the model to $2S$ 
equivalent orbitals on each site. 

The line shape of the obtained spectra in Fig. \ref{Erg_sc} is apparently 
distinctly different from the experimental data. As we argue in this paper 
the reason for this discrepancy is due to the neglect of resonance phenomena 
in the photon-magnon vertex function $V$; the non-resonant limit 
$(2\D -\w_i) \gg J$ for which the Loudon-Fleury theory is valid is 
not the relevant situation for the experimental Raman data  on cuprate 
antiferromagnets.

\section{Resonant Raman scattering}
\begin{figure}
\begin{center}
\epsfig{file=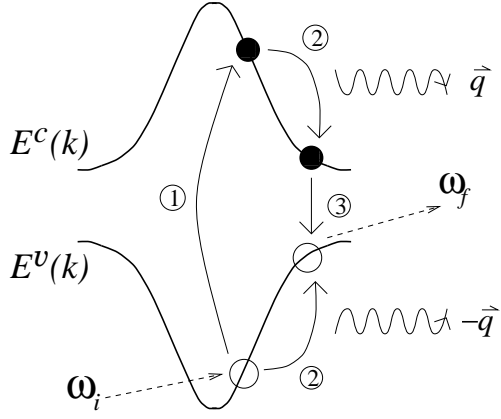,height=5.5cm, width=6.5cm}
\end{center}
\caption{\label{res-process}Scattering process which leads to  
the strongest resonant enhancement of the vertex function.
The dashed lines denote the incoming and outgoing photons, 
the wiggly lines denote the magnons. $E^v(\vk)$ and $E^c(\vk)$ mark the 
valence and conduction bands, respectively. 
$1.)$ Excitation of a particle-hole pair created by absorption 
of the incoming photon with frequency $\w_i$. $2.)$ Creation of a magnon 
pair. $3.)$ Recombination of the particle-hole pair and emission 
of the outgoing photon with frequency $\w_f$.}
\end{figure}

\begin{figure}
\begin{center}
\epsfig{file=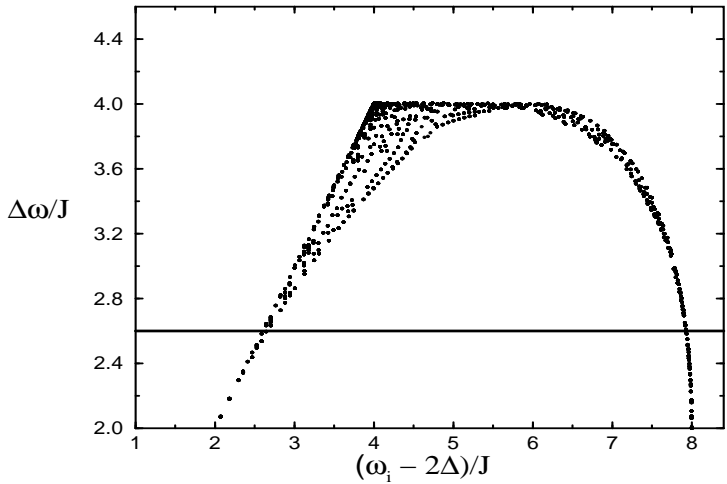, height=7.5cm, width=10.5cm}
\end{center}
\caption{\label{tpr-region}Numerical solution of the triple resonance 
          conditions. Each point in the figure corresponds to a pair 
          $(\w_i - 2\D,\dw)$ 
          where momenta $\vk$ and $\vq$ exist so that the resonance 
          conditions are simultaneously fulfilled. The solid line 
          marks the position of the two-magnon peak in the non-resonant 
          case. }
\end{figure}

In this chapter we consider the resonant case when the incoming 
photon frequency $\w_i$ is comparable to the SDW energy 
gap to within a typical magnon energy, $(\omega_i - 2\Delta )\sim J$. 
In this regime strong resonant enhancement of the vertex function becomes  
important due to the combination of photon induced interband transitions and  
the creation of a magnon pair. Resonances appear in all four vertex diagrams 
in Fig. \ref{vertices} which contribute to the total vertex function $V$. By 
inspection we find the strongest resonance to arise from a scattering 
process that is contained in the rhomb-shaped vertex diagram $A$
(Fig. \ref{vertices}a). 
Alternatively to Eq. (\ref{va}) this contribution to the vertex function 
is more  conveniently expressed in terms of SDW conduction and valence band 
quasi particle propagator matrices which allows more easily to identify the 
physical process responsible for the strongest  resonance. 
Explicitly we rewrite the algebraic result for the diagram $A$ as 
\bea
V_A({\vq},i\wm,i\w_t,i\w_p) & = & 
  \frac{1}{\beta}  \sum_{\On}\frac{1}{N}{\sum_{\vk}}^{\s}
               M_A(\vk ,\vq)\mbox{Tr}
    \left[ \right.  \Gg  ({\vk};\On )\nonumber\\ 
    &&\p (\vk ,\vk) \Gg ({\vk} ,\On +i\wm )
    \p (\vk ;\vk -\vq) \Gg ({\vk} -{\vq};\On +i\wm -i\w_p)\nonumber \\ 
    &&\p (\vk -\vq , \vk -\vq) 
  \Gg  ({\vk} -{\vq};\On +i\w_t -i\w_p )
    \p (\vk -\vq , \vk )
 \left.  \right]\qquad,
\label{vgg}
\eea
where we have introduced the coherence factor matrix 
\be
\p (\vk ,\vks )= \left( \begin{array}{cc} 
                 n_{\vk ,\vks } & m_{\vk ,\vks}  \\ 
                 m_{\vk ,\vks} & -n_{\vk ,\vks} 
    \end{array}\right)\quad 
\ee
with matrix elements 
\bea
n_{\vk, \vk -\vq}& = &u_{\vk}v_{\vk -\vq}-v_{\vk}u_{\vk -\vq} \qquad 
\nonumber \\
m_{\vk ,\vk -\vq}& = &u_{\vk}v_{\vk -\vq}+v_{\vk}u_{\vk -\vq} \qquad.
\eea
The symmetry factor $M_A(\vk ,\vq)$ depends on the scattering geometry 
and is defined in Eq. (\ref{sym2}) in the appendix.  
Carrying out the required matrix multiplications in Eq. (\ref{vgg}) leads to 
16 combinations of conduction $(G^c)$ and valence band propagators $(G^v)$. 
Among them we single out the term which leads to the by far strongest 
resonant enhancement, i.e. the term containing the product $G^vG^cG^cG^v$, 
and neglect in the following all other combinations 
with weaker resonances. The physical process underlying the dominant 
resonance term is shown in Fig. \ref{res-process}. 
The incoming photon creates a particle-hole pair by exciting an 
electron from the valence into the conduction band. This initial 
excitation then decays into a particle-hole pair with lower energy 
by the creation of two magnons with zero total momentum. Finally, the 
particle and the hole recombine under emission of the outgoing photon. 
After analytic continuation the vertex function contribution of 
the resonant scattering process shown in Fig. \ref{res-process} 
is given by 
\bea
V_A^R(\vq,\w_i ,\dw ) & = & \frac{1}{N}{\sum_{\vk}}^{\s}M_A({\vk},\vq )
\frac{m_{\vk ,\vk}}{(\w_i-2E(\vk) +i\d )}\times 
 \nonumber\\
&&\frac{n_{\vk, \vk -\vq}\quad m_{\vk -\vq
    ,\vk}\quad 
n_{\vk -\vq,\vkq}}{
(\w_i-\frac{\dw}{2}- E(\vk)-E(\vk-\vq) +i\d )(\w_i-\dw -2E(\vk-\vq)
+i\d )}\; .
\label{tpr-term}
\eea
Here we have already identified $\dw = \w _i - \w _f = 2\w_{sw} (\vq )$ 
as enforced by 
the energy conserving $\d$-functions contained in the imaginary parts of the 
transverse dynamic susceptibilities (see Eq. (\ref{chi_sc})). The strong 
resonant enhancement of the vertex part $V_A^R$ results from the 
possible simultaneous vanishing of its three energy denominators 
and has therefore been termed a "triple resonance" in Ref.\cite{Chubukov}.
A triple resonance occurs if the following conditions hold: 
\be 
\w_i = 2E(\vk ) \qquad , \qquad \w_i - \dw = 2E(\vk - \vq)\quad.
\label{cond1+2}
\ee
The triple resonance conditions have been studied in detail analytically in 
Ref.\cite{Chubukov} by Chubukov and Frenkel. We have confirmed their 
 results by solving numerically the triple resonance 
equations for $\w_i$ and $\dw $ for a sequence of magnon momenta $\vq$ in 
the MBZ.
The numerical results are shown in Fig. \ref{tpr-region} together with 
the position of the two-magnon peak in the non-resonant case. 
The boundary of the triple resonance region, i.e. the region in the 
$(\w_i,\dw)$ plane, where the resonance conditions can be matched, is 
determined by the high symmetry directions in the MBZ. Each point in the 
figure corresponds to a magnon momentum $\vq$ and marks the frequencies 
$\dw, \, \w_i$ for which the triple resonance equations (\ref{cond1+2}) 
can be solved for electronic momenta $\vk$ in the MBZ. The vertex 
function $V_A^R$ diverges for those values of $\w_i$ 
and $\D\w$, which lie in the indicated region (Fig. \ref{tpr-region}) 
(as long as quasi particle lifetime effects are neglected).

In our slightly different calculational scheme we have so far reconfirmed 
the results of Chubukov and Frenkel for the 
location of the divergency of the vertex function arising from the 
vanishing of the denominators in Eq. (\ref{tpr-term}).
The obvious important task is now to explore the change of the 
two-magnon Raman intensity profile due to the triple resonance vertex function.

In the non-resonant case we essentially exploited 
the $\g_{\vq}^d$ momentum dependence of the photon-magnon vertex 
function in the strong coupling limit for the calculation of 
the scattering intensity in $B_{1g}$ geometry. In this 
 case the $\g_{\vq}^d$ momentum dependence arises simply from the 
$B_{1g}$ symmetry factor $M_A(\vk ,\vq)$ and from neglecting the 
momentum dependence of the SDW quasi particle propagators
 to leading order in $t/U$. In the present case the symmetry factor is left 
unchanged but the structure of the resonant vertex function modifies the 
momentum dependence of $V_A(\vq)$.
Nevertheless, numerical calculations show that for intermediate photon 
frequencies $3<(\w_i-2\D)/J <5$ the $\vq$-dependence of 
$V_A^R$  in $B_{1g}$ geometry still follows very closely the $\gamma_{\vq}^d$ 
form. In order to solve the Bethe-Salpeter equation for the renormalized 
vertex $\Gamma$ even with a triple resonant photon-magnon vertex function 
we project $V_A^R$ on the $\g_{\vq}^d$-channel, i.e. we single out the 
$\gamma_{\vq}^d$ symmetry component of $ V_A^R$ by introducing   
\be
\tilde{V}_A^R(\vq,\w_i ,\dw )=g(\w_i
,\dw )\;\g_{\vq}^d \qquad,
\label{Proj}
\ee
where
\bea
g(\w_i , \dw )=\frac{\sum_{\vq}\s \gamma_{\vq}^d V_A^R(\vq,\w_i ,\dw )}
                    {\sum_{\vq}\s \left(\gamma_{\vq}^d\ \right)^2}\qquad.
\eea
Using the projection $\tilde{V}_A^R$ of the triple resonance vertex term 
instead of $ V_A^R$ the calculation of the $B_{1g}$ scattering intensity 
proceeds completely analogous 
to chapter VI  and leads to  
\be
I_{B_{1g}}(\dw)=\frac{1}{\pi}|g(\w_i ,\D\w)|^2\;\mbox{Im}\left\{\frac{L(\dw
    +i\delta)}{1-12J\,L(\dw +i\delta)}\right\}\, .
\label{tprerg}
\ee
The result for the resonant $B_{1g}$ Raman intensity thus factorizes into 
the absolute square of the $\g_{\vq}^d$ symmetry component of the vertex 
function and the two-magnon part which remains unchanged.
The spectrum therefore consists of two separate contributions of distinct 
origin:
the two-magnon peak at $\D\w\approx 2.6J$ and the triple resonance peak 
which appears well above the two-magnon peak close below $4J$ 
for an intermediate photon frequency range $3<(\w_i -2 \D)/J <5$ 
(see Fig. \ref{tpr-region}).

Clearly, the magnitude of the triple resonance structure in the two-magnon 
intensity profile depends on the strength of the quasi particle damping 
 resulting from self-energy corrections due to residual interactions 
between the SDW quasi particles \cite{oldi}. Similarly, its precise location 
depends on the renormalized band dispersion. Here, for 
the purpose of demonstrating the consequence of the triple resonance vertex 
on the two-magnon line shape, we model the effect of quasi particle damping 
by adding a finite imaginary part to the energy 
denominators  of the triple resonant vertex function. For the choice of 
a typical quasi particle lifetime we are guided by the results of a 
self-consistent non-crossing calculation of the self-energy correction 
in the SDW state of the half-filled Hubbard model \cite{oldi}.
The result for the  two-magnon  
intensity in $B_{1g}$ geometry with magnon-magnon interactions and a 
broadened triple resonance vertex function is shown in  Fig. \ref{int-fig}.

\section{Discussion}
\begin{figure}
\unitlength1.5cm
\begin{picture}(8.5,10.2)(0,0)
\put(0.4,-10.5){
\epsfig{file=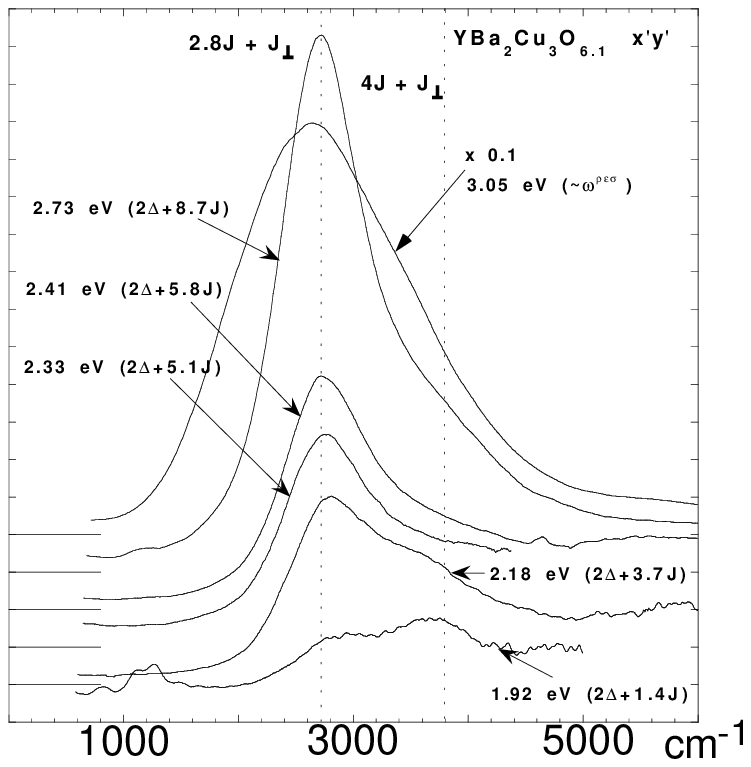,height=27.9cm}}
\put(-0.2,5.2){ 
 \epsfig{file=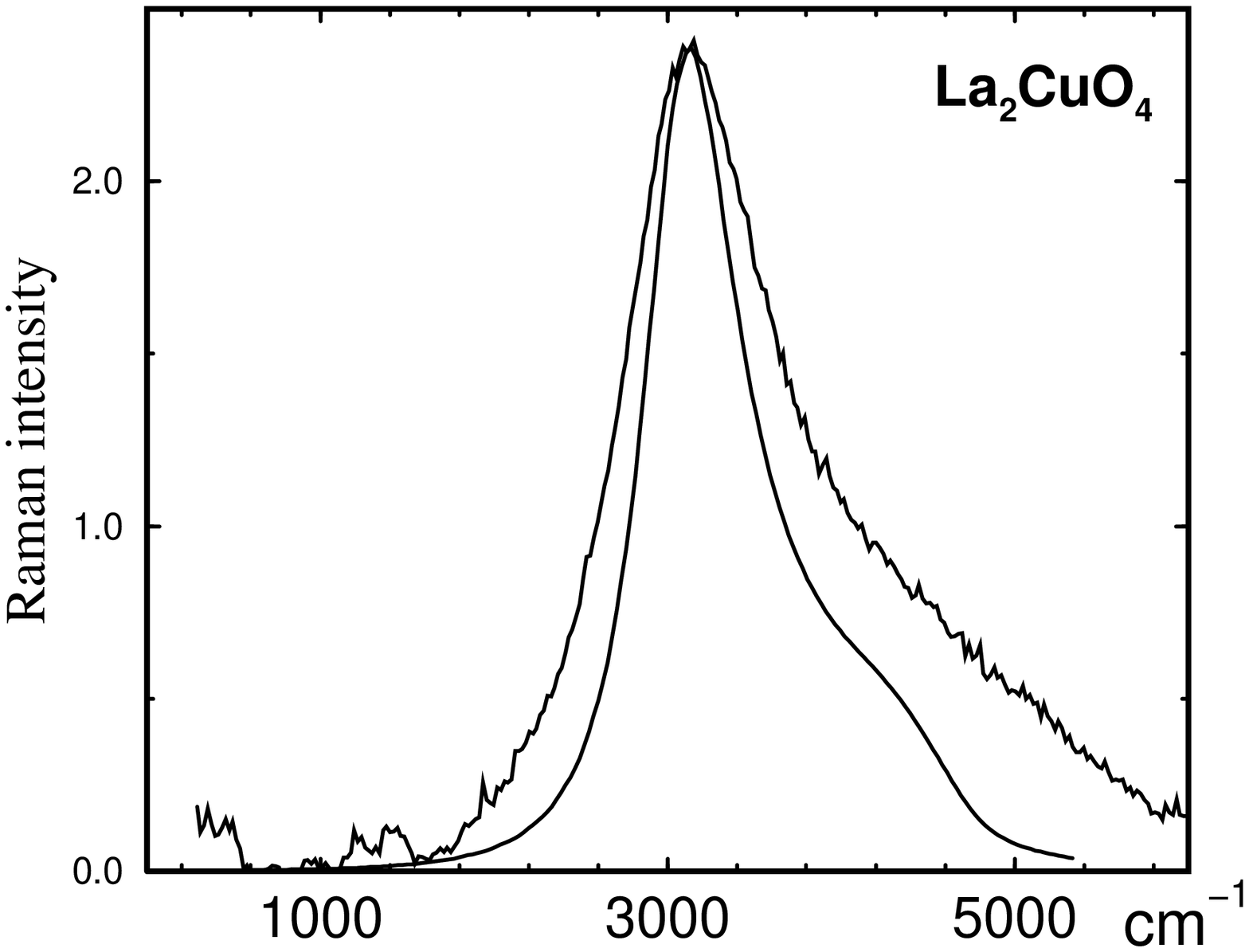,height=8.181cm}} 
\put(0,10){$a)$}
\put(0,5){$b)$}
\put(4.5,0.2){\bf\large$\dw$}
\put(4.5,6.4){\bf\large$\dw$}
\end{picture}
\caption{\label{int-fig}$a)$ Raman scattering intensity in arbitrary units 
in $B_{1g}$ geometry as calculated  from  Eq. (\ref{tprerg}) with 
$(\w_i -2\D )= 3.6J$ for a lattice with $100\times 100$ sites. For 
the two-magnon term an imaginary broadening of $i\d =i0.09J$ was used 
in the frequency denominators, while the vertex part was evaluated with 
$\d =0.4J$. The jagged line displays the experimental spectrum 
of $La_2CuO_4$ taken at room temperature with a laser frequency of 
$\w_i = 2.55 eV$ \protect\cite{Lyons}. As in Ref. \protect\cite{Singh} 
a background was substracted from the Raman data. 
For comparison with the data the magnetic energy scale was set 
to $J = 1200 cm^{-1}$ and the peak intensities were scaled to 
coincide. 
$b)$ Experimental spectra of $YBa_2Cu_3O_{6.1}$ at $T= 5K$ for various  
laser energies as indicated in the figure 
(reproduced from Ref.\protect\cite{Blumberg}).}
\end{figure}

Fig. \ref{int-fig} shows a comparison of our result for the two-magnon 
Raman intensity in $B_{1g}$ geometry to the experimental
 spectrum for $La_2CuO_4$ taken from Ref.\cite{Lyons}. 
In order to isolate the two-magnon signal a background was substracted 
from the experimental data as in Ref.\cite{Singh}.
The calculated lineshape 
is dominated by the two-magnon peak but its high energy shoulder  results 
 solely from the triple resonance enhancement in  the photon-magnon  
vertex function. The charge transfer energy gap in $La_2CuO_4$ is about 
$2eV$ as deduced from measurements of the optical conductivity \cite{Uchida}. 
So for the laser photon frequency $\w_i =2.55 eV$ used in the Raman 
experiment we have $\w_i - 2\D\sim 550meV$. 
For the calculation of the Raman intensity we chose $J=1200 cm ^{-1}$ 
such that the two-magnon peak frequency coincides with the experimental 
value and $\w_i - 2\D = 3.6 J\sim 550 meV$.  
For intermediate photon frequencies $(\w_i -2\D)\sim 4J$ we obtain a 
 lineshape that is in fair agreement with the experimental spectra 
not only for $La_2CuO_4$ but also for the AF double-layer compound 
$YBa_2Cu_3O_{6.1}$ near resonance as shown in Fig. \ref{int-fig}b 
for a series of photon frequencies \cite{Blumberg}.

Besides the high energy shoulder other experimental features in the 
scattering intensity  find a natural explanation as well by the interplay 
between the two-magnon peak and the triple resonance: 
\newline
$(a)$ $Two$ $separate$ $resonance$ $frequencies.$ Recent measurements of 
the Raman intensity by Blumberg et al. \cite{Blumberg} show an enhancement of 
the absolute Raman intensity for two different, well separated photon 
frequencies $\w_i$. 
In our analysis the Raman intensity as a function of $\w_i$ is expected to 
increase if the frequencies for the two-magnon and the triple resonance peak 
 merge. This does, in fact, happen for two distinct frequencies 
$(\w_i -2\D )\approx 2.6 J$ and $(\w_i -2\D) \approx 8J$. These 
frequencies can be read off from Fig. \ref{tpr-region} from the 
intersections of the two-magnon peak frequency (horizontal line) 
with the triple resonance region.\newline
$(b)$ $\w_i^{res} > 2\D.$ The same arguments as in $(a)$ apply also for 
understanding the experimental observation that resonance occurs for incoming 
photon frequencies well above the insulating energy 
gap $2\D$. This has already been 
pointed out previously in Ref.\cite{Blumberg}.
\newline
$(c)$ $Changing$ $line$ $shape.$ The experimental spectra in 
Fig. \ref{int-fig}b show a 
decreasing asymmetry with increasing photon frequency $\w_i$. The two 
spectra for $YBa_2Cu_3O_{6.1}$ taken at the lowest frequencies contain two well separated contributions, 
whereas the total line shape of the spectrum of the largest laser frequency 
is nearly symmetric and the two peaks are indistinguishable. 
(For a detailed discussion of these spectra see Ref.\cite{Sugai}.)
Within our calculation, the almost symmetric intensity 
profile corresponds to the near coincidence of the triple resonance and 
the two-magnon peak frequencies. 

Fig. \ref{int-fig} a shows  Raman data of $La_2CuO_4$, which were taken at 
room temperature \cite{Lyons}. The most noticable difference to the calculated 
intensity is the larger peak width in the experimental spectrum. 
There are several possible explanations for this discrepancy.
First of all one might wonder about the influence of finite temperatures, 
since our calculation was performed at $T=0$. However, a comparison between 
the low and high temperature ($T\sim T_{N\acute{e}el}$) spectra in Fig. 
\ref{int-fig}a and Fig. \ref{int-fig}b for $La_2CuO_4$ and 
$YBa_2Cu_3O_{6.1}$,  respectively, reveals that the thermal effects are 
of minor importance since the spectra have a comparable peak width.  
Clearly, lifetime effects due to magnon-phonon and also magnon-magnon 
interactions in particular for the zone boundary magnons  will broaden 
the two-magnon Raman signal. The Raman spectrum in Fig. \ref{int-fig}a 
was evaluated on a $100 \times 100$ lattice and a small broadening of 
$i\d = i 0.09J$ in the energy denominator of the dynamic spin susceptibility 
was introduced only for calculational purposes.  
The effects of magnon-phonon and magnon-magnon interactions were neglected 
in the present work and the sharper peak  of the calculated Raman intensity in 
comparison  to the data is expected on physical grounds.
The intention of our present analysis was, instead, to show that the resonant 
frequency dependence of the vertex function for the photon-magnon  
coupling - similar to and in agreement with the recent work by Chubukov 
and Frenkel \cite{Chubukov} - leads to a qualitative change of the two-magnon 
line shape and provides a natural explanation for the resonance phenomena 
observed in light scattering experiments on cuprate antiferromagnets.

Furthermore we neglected the contribution from four-magnon and higher 
order scattering processes. Canali and Girvin have attempted to include the 
four-magnon contribution in the framework of the Heisenberg model and the 
Loudon-Fleury coupling Hamiltonian \cite{Canali}. Although the first three 
moments of the resulting scattering intensity are in good agreement with 
experiment and also with the series expansion results by Singh et al. 
\cite{Singh}, the experimental line shape is not successfully explained. 
The integrated four-magnon intensity is only 2.9\% of the two-magnon 
intensity, and if this estimate is correct, the four-magnon contribution 
will barely affect the overall Raman intensity profile. Furthermore, 
the Canali and Girvin four-magnon peak frequency is 2.5 
times larger than the two-magnon peak frequency and therefore one has to 
conclude that four-magnon scattering cannot be responsible for the structure 
in the Raman intensity near $4000cm^{-1}$.

The effects of  phonon-magnon interaction have been studied 
previously in Refs.\cite{Knoll,Sanger,Nori} for Raman scattering in the 
Heisenberg antiferromagnet. As expected, the damping of 
zone boundary magnons does indeed lead to a significant 
broadening of the two-magnon signal. A qualitatively smaller 
contribution to the damping arises also from magnon-magnon interactions \cite
{Kopietz}.  
Nori et al. considered the effect of a random Gaussian variation
$\d J_{ij}$ of the exchange coupling $J_0$ assumed to originate 
from a distortion of the crystal lattice due to low frequency lattice 
vibrations \cite{Nori}. Using numerical 
techniques Nori et al. showed that in this model a broad and asymmetric 
two-magnon line shape with an enhancement of spectral weight at 
higher energies emerges if the mean deviation $<\d J_{ij}> $ is as 
large as $J_0/2$. 
Such an enormous variation of the exchange coupling, however, appears 
physically unreasonable. It is far from being clear whether 
more moderate values of $<\d J_{ij}> $ are sufficient for this mechanism 
to explain the line shape of the two-magnon Raman spectra. 
We emphasize that in all these works 
\cite{Knoll,Sanger,Nori} the resonant nature of the two-magnon 
scattering process was not taken into account.

\section{Summary and Conclusions}
Based on the single-band Hubbard model at half filling we have performed 
a microscopic analysis of two-magnon Raman scattering in a SDW 
antiferromagnet. 
In a diagrammatic formulation we have explicitly taken into account the 
structure of the photon-magnon coupling vertex. This allows to explore both,
the non-resonant ($\w_i\ll 2\D$) and the resonant case ($\w_i -2\D\sim J$) 
for the light scattering intensity, where the latter is the relevant limit 
for Raman experiments on undoped cuprate antiferromagnets.
For the non-resonant case in the strong coupling limit the results of the 
conventional Loudon-Fleury theory for two-magnon Raman scattering in the
Heisenberg antiferromagnet are almost quantitatively reproduced verifying 
our selection of the relevant vertex diagrams.
In the resonant regime we identified the physical process which yields the 
strongest diverging part of the 
photon-magnon  vertex function 
and analyzed the conditions and consequences of a triple resonance.
For photon frequencies well above the SDW energy gap triple resonance occurs 
for transferred photon frequencies larger than the two-magnon peak frequency 
leading to a high energy shoulder of the two-magnon intensity profile. 
These results confirm the conclusions of the triple resonance theory by 
Chubukov and Frenkel \cite{Chubukov} for resonant two-magnon Raman scattering. 
In this theory effective electron-magnon vertices were constructed from the 
requirement that these vertices lead to the strong coupling $RPA$ form of 
the transverse dynamic spin susceptibility. In contrast to the work of 
Chubukov and Frenkel we have performed a more systematic microscopic 
derivation for the two-magnon scattering intensity and we have evaluated 
the intensity profile by explicitly taking into account the resonant 
structure of the photon-magnon coupling vertex.

By using unrenormalized SDW quasi particle propagators the triple resonance 
leads to a true divergence of the vertex function. Self-energy corrections, 
however,
 will remove the divergence leaving a finite enhancement in the Raman 
intensity at the triple resonance frequency. These self-energy  
corrections have been calculated in previous work \cite{oldi}. We have 
here for simplicity modelled quasi particle lifetimes effects by adding 
a constant damping term to the propagators in the triple resonance vertex 
function. A more reliable quantitative 
estimate for the strength of the triple resonant enhancement will 
require a refined treatment with the inclusion of self-energy corrections.

In conclusion, we found that the resonant frequency dependence of the 
photon-magnon vertex function gives rise to an enhancement of the high 
energy side of the two-magnon Raman peak in $B_{1g}$ scattering geometry. 
The combination of resonant transitions between the SDW quasi particle 
bands and magnon pair excitations provides a microscopic basis for 
understanding the resonant Raman scattering experiments on cuprate 
antiferromagnets.

\section*{Acknowledgments}
This work has been performed within the program of the Sonderforschungsbereich 
341 supported by the Deutsche Forschungsgemeinschaft (DFG). A.P. Kampf 
gratefully acknowledges the support through a Heisenberg fellowship of the 
DFG. We thank A.V. Chubukov, and W. Brenig for stimulating discussions and 
 correspondence and K.B. Lyons for providing us with some of his Raman 
data files.

\begin{appendix}
\section*{}
For the calculation of the vertex diagrams shown in Fig. \ref{vertices} 
we use the bare SDW single particle propagator ${\bf G}_0^{\si}$ 
(in $c$-fermion representation) in $2\times 2$ matrix notation. In the finite 
tempe\-rature Matsubara formalism the vertex functions of 
the four diagrams are given explicitly by
\bea
  V^\si _A({\vq},i\wm,i\wt,i\wp)&=&
  \frac{1}{\beta}\sum_{\On}\frac{1}{N}{\sum_{\vk}}^{\s}
  M_A({\vk},\vq )\mbox{Tr}\left[\Gs({\vk};\On ) 
 \sz \,\Gs ({\vk} ;\On +i\wm ) \right. \nonumber\\ &&
     \Gms ({\vk} -{\vq};\On +i\wm-i\wp) 
\left. \sz \;\Gms ({\vk} -{\vq};\On +i\wt -i\wp
  )\right]\; ,
\label{va}
\\
\nonumber\\
V^\si _B({\vq},i\wm,i\wt,i\wp)&=&\frac{1}{\beta}\sum_{\On ,\tOn}
\frac{1}{N}{\sum_{{\vk},{\vl}}}^{\s} M_B({\vk})
  \mbox{Tr}\left[\Gms({\vk};\On)\sz 
                   \Gms({\vk};\On+i\wm )\nonumber \right.\\
 & &\left.\sz\Gms({\vk}-{\vq};\On +i\wt )\right]\,U\,
\mbox{Tr}\left[\Gs({\vl};\tOn)\nonumber \right.\\
 & &\left.\Gs({\vl};\tOn+i\wt )
\Gms({\vl}-{\vq};\tOn +i\wt -i\wp )\right]\; ,
\label{vab}
\\
\nonumber\\
V^\si _C({\vq},i\wm,i\wt,i\wp)&=&
\frac{1}{\beta}\sum_{\On}\frac{1}{N}{\sum_{\vk}}^{\s}
  M_B({\vk})\mbox{Tr}\left[\Gs({\vk};\On)\sz \;
    \Gs({\vk};\On+i\wm)\sz \;\right.\nonumber\\
& & \left.  \Gs({\vk};\On +i\wt)\Gms({\vk}-{\vq};\On +i\wt-i\wp)
\right]\; ,
\label{vc}
\\
\nonumber\\
V^\si _D({\vq},i\wm,i\wt,i\wp)&=&
\frac{1}{\beta}\sum_{\On}\frac{1}{N}{\sum_{\vk}}^{\s}
 M_B({\vk})
 \mbox{Tr}\left[\Gs({\vk};\On)\sz \;
        \Gs({\vk};\On -i\wm)\sz \right.\nonumber\\
& &\left. \Gs({\vk};\On+i\wt)\Gms({\vk}-{\vq};\On +i\wt-i\wp)\right]\; .
\label{vd}
\eea
The symmetry factors of the basic scattering vertices from the coupling of 
the photon vector potential to the electron current density have 
been combined into the functions
\be
M_A(\vk ,\vq)=\frac{2\pi e^2}{\sqrt{\omega_{i}\omega_{f}}\:\Omega}\sum_{\alpha\beta}
\frac{\partial \e_{\vk}}{\partial k_{\alpha}}\frac{\partial \e_{\vk +\vq}}
{\partial k_{\beta}}e_i^{\alpha}e_f^{\beta}\qquad,
\label{sym2}
\ee
\be 
M_B(\vk )=M_A(\vk ,\vec{0})
\qquad.
\label{symFac}
\ee
Note that the factor ${\partial \e_{\vk}}/{\partial k_{\alpha}}$ 
changes sign when umklapp scattering $(\vk\rightarrow\vk +\vQ)$ has 
taken place along the fermion lines. In the $2\times2$ matrix formulation 
the alteration of the sign is conveniently taken 
into account by inserting the Pauli matrix $\sz$.
Including all diagrams with reversed direction of the fermion lines 
yields a factor two in the total vertex function.

\end{appendix}

\end{document}